\documentclass{aa}
\usepackage{amsmath}
\usepackage{natbib}
\usepackage{graphicx}
\bibpunct{(}{)}{;}{a}{}{,} 

\begin{document}

   \title{Detection of 25 new rotating radio transients at 111 MHz}

   \author{
          S.A. Tyul'bashev
                    \and
         V.S. Tyul'bashev
                  \and
          V.M. Malofeev}
   \institute{
Lebedev Physical Institute, Astro Space Center, Pushchino Radio Astronomy Observatory,\\
142290, Moscow region, Russia\\
  \email{serg@prao.ru}
          }
   \date{Received , 2018; accepted , 2018}

  \abstract
  {Nearly all fast radio RRAT-type transients that are pulsars with rare pulses have been previously detected using decimetre wavelengths.
We present here 34 transients detected at metre wavelengths in our daily monitoring at declinations $-9\degr \le \delta \le +42\degr$.
 25 transients are new RRATs. We confirmed the detection of 7 RRATs based on our early observations. One of the 34 detected transients was determined to be a new pulsar J1326+3346. At the same time, out of 35 RRATs detected at the decimetre wavelengths and included in the studied area, only one was detected by us J1848+1518. The periods of 6 RRATs were found from the time of arrival of single pulses.
Three quarters of all RRATs were observed more than once and the total number of RRATs in the area studied has doubled.}

\keywords{<pulsars - rotating radio transients (RRATs)>}

      \maketitle
%

\section{Introduction}

  The detection of fast radio transients in archival records of the Parkes Pulsar Survey resulted in the discovery of two classes of objects.
   The first group, rotating radio transient (RRAT), was determined to consist of galactic objects. These are pulsars with rare pulses (with intervals between pulses
 in the range of minutes or hours) \citep{McLaughlin2006}.
   The second group was fast radio burst (FRB), which are rarer objects of an extragalactic nature \citep{Lorimer2007}.

   The duration of flares of the RRATs and FRB objects is comparable; the time scale is from a fraction of a millisecond to tens of milliseconds,
   while the observed dispersion measures ($DM$) appreciably differ (see catalogues ATNF {\it http://www.atnf.csiro.au/people/pulsar/psrcat/} \citep{Manchester2005},
   RRATalog {\it http://astro.phys.wvu.edu/rratalog/},
   FRBcat {\it http:// www.astronomy.swin.edu.au/pulsar/frbcat/} \citep{Petroff2016}).
   Whereas the dispersion measures'
 intervals for known pulsars, $2.38 < DM < 1778$ pc/cm$^{3}$ (ATNF), and for RRATs, $9.2 < DM < 786$ pc/cm$^3$ (RRATalog),
   are rather similar, this interval is offset for FRB toward higher values, $176.4 < DM < 1629.18$ pc/cm$^3$ (FRBcat).
   For distinguishing RRATs, one examines the first turn for repeated pulses and searches for the pulsar period.
   If the period cannot be determined, the separation of the RRAT and FRB can be achieved using the model of the distribution of thermal electrons in the Galaxy \citep{Cordes2002,Yao2017}
   and compares in the case of the new source, the observed dispersion measure with the one expected in that direction.
   If the observed value is considerably higher than expected, then one tells about the detection of FRB.
   In fact, from the point of view of the observer, the search for fast radio transients, either RRATs or FRB, is essentially a search for single pulses from pulsars.
   If we recall history, the discovery itself of pulsars as a new class of objects took place after the detection of their single pulses \citep{Hewish1968}.

   So far, fast transients (RRATs/FRBs) have been found mostly in the decimetre wavelength range using the largest telescopes in the world, 64-metre Parkes, 100-metre Green Bank, 300-metre Arecibo
   (see, e.g. \citet{Keane2010,Karako2015,Deneva2016}).
   Dedicated searches for RRATs/FRBs were also carried out in the metre wavelength range at LOFAR (LOw Frequency ARray), but the result was negative \citep{Karastergiou2015}.
   Meanwhile, the sensitivity of LOFAR and higher frequency surveys are comparable.
   RRAT detection at metre wavelengths was used for the one discovered in Pushchino at the Large Phased Array radio telescope \citep{Shitov2009} before starting new daily monitoring survey.

   The optimal combination of conditions for the search of fast transients inlcude:
   high instantaneous sensitivity of the radio telescope, because fast transients are, as a whole, rather faint objects;
   short sampling interval comparable with the pulse duration;
   multichannel receiver with a narrow band of each frequency channel to reduce pulse broadening due to the dispersion measure.
   These conditions are partly satisfied in the monitoring survey initiated in 2013 at the Large Phased Array of the Lebedev Physical Institute (LPA LPI) at a frequency of 111 MHz.
   The aim was to forecast the time of arrival of coronal mass ejections from the Sun on the basis of observations of scintillating radio sources \citep{Shishov2016} (Space Weather Project).
   The survey covered an area of about $17000$~square degrees daily.
   The data from the survey were also used for the search of pulsars and transients.
   Two independent groups have discovered 41 pulsars \citep{Tyulbashev2016,Rodin2017,Tyulbashev2017}.

   In 2017 a test search for RRATs in daily monitoring data was also performed using 5 days (1-5 September 2015)\citep{Tyulbashev2017a} and 28 days (1-28 September 2015) \citep{Tyulbashev2018}.
   As a result, 54 fast transients have been detected; of these, seven RRATs were absent in ATNF and RRATalog, and 47 have been identified with known pulsars.
   In this paper, we present the results of our search for new fast transients after processing six months of monitoring data.

\section{Observations and results}

   Our monitoring observations were conducted using 96 beams simultaneously covering about $40$~square degrees in the sky at the LPA LPI meridian telescope at a frequency of $111$~MHz and commenced from August 2014.
   The beams covered the declination interval $-9^{\circ} < \delta < +42^{\circ}$ and the beam size was $\sim 0.5^{\circ} \times 1^{\circ}$.
   The data were recorded in one-hour segments.
   Each segment was initialised by GPS time markers, but within the hour time was controlled by a quartz oscillator.
   Transients were searched for in the data recorded in a 2.5-MHz band in a 32-channel frequency mode with a sampling interval of $12.5$~ms.
   In this mode, the fluctuation sensitivity of the telescope with the effective area of the antenna $45000$~m$^2$ toward zenith is $0.3$~Jy.
   The data from July-December 2015 have been processed.
    The detailed technique of the  search for the transients was described in the paper \citet{Tyulbashev2018}. The main approach was that the data in 32-frequency channels were summed for each DM, every peak in the mean profile was registered, and  a few criteria were taken into account.

   The software performed searches for dispersion measures from $3$ to $100$~pc/cm$^3$ and searched for dispersed signals with a signal-to-noise ratio (S/N) $> 5.5$.
    A limit of the DM was taken because of the large number of false sources that exists for DM less than $3$~pc/cm$^3$ and we did not have any detections in our pulsar search for $DM>100$~pc/cm$^3$ \citet{Tyulbashev2016,Tyulbashev2017}.

   After the primary search, the catalog included about 300000 objects, which were analysed by means of the BSA-Analytics software ($Qt/C++$ code distributed under the $GPL\, V3.0$ license.
   See {\it https://github.com/ vtyulb/BSA-analytics)}.
   The BSA-Analytics programme during the search for transients gave the choice of a certain beam (sky direction), restrictions of the signals detected at a given signal-to-noise ratio,
   and the sidereal time of the tested block.
   The filters of the programme checked whether the transient was detected in no more than three beams,
   or detected only once and at only one dispersion measure,
   or the main part of pixels with a large signal-to-noise ratio are arranged along a certain dispersion measure (geometric filter),
   or removal of poor-quality days, or elimination from the shown catalogue of the objects already found earlier.
   The software also enables us to search for the RRAT period, if during the observational sessions two or more pulses per session were detected,
   to add together the dynamic spectra and mean profiles,
   to determine the pulse halfwidth, to save graphs,
   and to refine the dispersion measure by overlaying a dispersion curve with the known dispersion measure on the dynamic spectrum.
   This software also lends some other possibilities of usage.
   The ultimate control of the identified candidates was done visually for transient with visible line of dynamic spectra. This visible line corresponds to the value S/N>7 of pulse profile.

   Additional details of the capabilities of the updated LPA LPI and of the search technique are given in \citet{Shishov2016,Tyulbashev2017,Tyulbashev2018}
   and in BSA-Analytics Project ({\it https://bsa-analytics.prao.ru/}).

Figure 1 demonstrates the sample pulses from the pulsar and the new RRAT. We consider the objects J0941+1621 and J0943+1631 separately.
J0943+1631 is a known pulsar with a dispersion measure of $20.3$~pc/cm$^3$, and the overwhelming part of the dispersed signals with a dispersion measure of $19-20$~pc/cm$^3$ is visible in this direction.
For some of these records, another transient appears, which can be at an arbitrary place in the dynamic spectrum graph with respect to the pulsar, and it has a different dispersion measure.
For better visibility, we averaged 19 dynamic spectra and mean pulsar profiles. The red line in the averaged dynamic spectrum of the left part of the image corresponds to the dispersion measure of $19$~pc/cm$^3$.
In the figure with the J0941+1621 RRAT a red line is also drawn, but it corresponds to a dispersion measure of $24$~pc/cm$^3$.
The graphs for the RRAT J0941+1621 and pulsar J0943+1631 are placed side by side to explicitly show the
different slopes of the dispersion curve of these two objects located in the same direction and having similar dispersion measures.
The average profile of the figure is the sum of two different dispersion curves and therefore, it has a complex shape with a triple peak.
One pixel visible on a dynamic spectrum picture has the time scale (12.5 ms) in the horizontal direction and a frequency channel bandwidth (78 kHz) in the vertical direction. The upper and lowest lines of the spectra correspond to the frequencies 111.461 and  109.039 MHz respectively, therefore, the dark line on the dynamic spectrum tracing the transient goes from the top of the right corner to the bottom of the left corner.
The dynamic spectrum is presented with a small margin in time. This time in milliseconds is shown under the transient name.

  \begin{figure}
   \centering
   \includegraphics[width=\columnwidth]{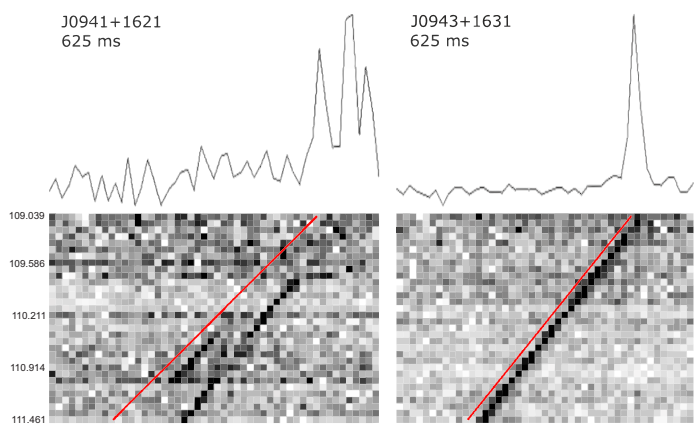}
     \caption{There are two different slopes on the dynamic spectrum in the image which corresponds to RRAT J0941+1621 and PSR J0943+1631.}
   \end{figure}

Table 1 lists the main parameters of the new 25 RRATs together with additional information for eight RRATs published earlier by \citet{Samodurov2017,Tyulbashev2018} and in RRATalog.
The first column contains the RRAT's name in the J2000 notation.
The second and third columns give the right ascension and declination of the source.
The fourth column lists the estimated dispersion measure of the RRAT.
In the fifth column, estimates of the observed peak flux density $S_{peak}$ for the strongest pulses
are given with corrections for the declination (zenith correction) and for the particulars of the multibeam LPA LPI directivity pattern.
In the sixth column, the observed pulse halfwidth is given without taking into account dispersion broadening within the channel frequency band.
The seventh column lists estimation of RRAT periods.
These estimations were made for RRATs with a few detections of the individual pulses during one session. We measure the time distance between pulses and therefore can calculate the upper limit of the period.
In the eighth column, the number of repeated events is given.
In the ninth column, the references to the early detections of 8 RRATs are given.
For a single detection of a RRAT, the accuracy of the coordinates in the right ascension is $\pm 2^m$.
For multiple detections, the column lists the average coordinate and the rms error.
The declination accuracy is indicated in the table and the accuracy of the estimated period is $\pm 0.003$~s.
The flux density was estimated from the noise trace; therefore, the accuracy of this quantity is low.
It may differ from the given value by a factor of $1.5-2$.
For the RRATs with an estimated value for the period, we checked the presence of the signal
in the cumulative Fourier power spectra obtained for six-channel frequency data with a sampling interval of $0.1$~s \citep{Tyulbashev2017}.
We have found no harmonics for the periods listed in the table.
We have also tried averaging with the known period and dispersion measure for the days when multiple transients pulses were observed, taking into consideration that we might have detected flaring pulsars.
In the process of averaging over an observational session, we have not managed to obtain mean profiles.
\begin{table*}
\caption{Main parameters of 33 detected RRATs}
\label{table:1}
\begin{tabular}{lllllllll}
\hline
Transient name & $\alpha_{2000}$ &$\delta_{2000}$ & $DM $          & $S_{peak}     $ & $W_e $     &	$P   $ &	N  & ref. \\
               &                 &                & (pc/cm$^3$)     & (Jy)          & (ms)        &	(s)  & (times)& \\
\hline
J0139+3310     & $01^h39^m30^s \pm 20^s$ &$33^o10^\prime \pm 15^\prime$ & $20\pm 1$      &	40           &	25	  &     1.2473 &	$>10$ & [1] \\
J0305+4001     & 03 05 26 $\pm$ 32 & 40 01 $\pm$ 15 & $24\pm 2$	   &    2.7	     & $<12.5$    &	-      &	2     & \\
J0318+1341     & 03 18 35 $\pm$ 28 & 13 41 $\pm$ 20 & $12\pm 1$      &	7.6	     &  30        &	1.9740 &        $>10$ & [2] \\
J0452+1651     & 04 52 56 $\pm$ 90 & 16 51 $\pm$ 25 & $19\pm 3$      &	3.1	     & $<12.5$    &	-      &	2     & \\
J0534+3407     & 05 34 30 $\pm$ 20 & 34 07 $\pm$ 15 & $24.5\pm 1.5$  &	2.6	     &   15	  &     -      &	7      & \\
J0609+1635     & 06 09 18 $\pm$ 21 & 16 35 $\pm$ 30 & $85\pm 3$	   &    2.3	     &   55	  &     -      &	$>10$  & \\
J0625+1730     & 06 25 19 $\pm$ 90 & 17 30 $\pm$ 25 & $58\pm 4$	   &    4.3	     &   25	  &     -      &	3      & \\
J0640+0744     & 06 40 40 $\pm$ 35 & 07 44 $\pm$ 25 & $52\pm 3$      &	8.7	     &   35  	  &     -      & 	$>10$ & [2]\\
J0803+3410     & 08 03 05 $\pm$ 85 & 34 10 $\pm$ 15 & $34\pm 2$	   &    3.4	     &   25	  &     -      &	5     & \\
J0941+1621     &  09 41 30        &  16 21 $\pm$ 20 & 23-24          &	-            &   ??       &     -      &        $>10$ & \\
J1005+3015     & 10 05 30 $\pm$ 40 & 30 15 $\pm$ 15 & $17.5\pm 1.5$  &    28.3         & 	 30	  &     -      & 	$>10$ & [1] \\
J1132+0921     & 11 32 00         &  09 21 $\pm$ 20 & $22\pm 2$	   &     7.3	     &   40	  &     -      &	1     & \\
J1132+2515     & 11 32 50 $\pm$ 30 & 25 15 $\pm$ 15 & $23\pm 3$      &     6.2         &   20	  &     1.0020 &	$>10$ & \\
J1329+1349     & 13 29 00         &  13 49 $\pm$ 20 & $12\pm 2$      &	11.3         &	15        &	-      &	1     & [2]\\
J1336+3346     & 13 36 26 $\pm$ 50 & 33 46 $\pm$ 15 & $8\pm  1$      &	8.4          &	15        &	3.0130 &	$>10$ & [2]\\
J1346+0622     & 13 46 00	       & 06 22 $\pm$ 25 & $8\pm  1$      &	11.6         & $<12.5$    &	-      &	1     & \\
J1400+2127     & 14 00 18 $\pm$ 30 & 21 27 $\pm$ 15 & $10.5\pm 1$	   &    24.5         &	25        &	-      &	9  &   \\
J1404+1210     & 14 04 49 $\pm$ 38 & 12 10 $\pm$ 20 & $17\pm 1$      &	10.4         &	-	  &     2.6505 &	$>10$ & \\
J1502+2813     & 15 02 09 $\pm$ 40 & 28 13 $\pm$ 15 & $14\pm 1.5$    &	16.3	     &  25        &	3.7840 &	$>10$ & \\
J1555+0108     & 15 55 58 $\pm$ 17 & 01 08 $\pm$ 25 & $18.5\pm 1.5$  &	22.6         &	20        &	-      &	3     & [2]\\
J1732+2700     & 17 32 24 $\pm$ 41 & 27 00 $\pm$ 15 & $36.5\pm 1.5$  &	5.9          &	25        &	-      &	7     & \\
J1841-0448     & 18 41 10 $\pm$ 20 & -04 48 $\pm$ 25& $29\pm 3$      &  	3.8          &	15        &	-      &        $>10$ & \\
J1848+1518     & 18 48 42 $\pm$ 90 & 15 18 $\pm$ 20 & $75\pm 5$      &	2.3          &	25        &	-      &	2     & [3]\\
J1917+1723     & 19 17 30         &  17 23 $\pm$ 20 & $38\pm 3$      &	2.8          &	25        &	-      &	1     & \\
J1930+0104     & 19 30 30         &  01 04 $\pm$ 25 & $42\pm 3$      &	4.9          &  30        &	-      &        1  &    \\
J2052+1308     & 20 52 21 $\pm$ 40 & 13 08 $\pm$ 20 & $42\pm 3$      &	2.9          &	20        &	-      &	3     & \\
J2105+1917     & 21 05 20 $\pm$ 20 & 19 21 $\pm$ 20 & $33\pm 3$      &	2.7          &	25        &	-      &	10    & \\
J2107+2606     & 21 07 30         &  26 06 $\pm$ 20 & $10.5\pm 2$    &	3.1          &	25	  &     -      &	1     & \\
J2135+3032     & 21 35 00         &  30 32 $\pm$ 15 & $63\pm 2$      &	3.5          &	50        &	-      &	1     & \\
J2146+2148     & 21 46 00         & 21 48 $\pm$ 20 & $43\pm 3$      &	2.5          &	20        &	-      &	1     & \\
J2202+2147     & 22 02 21 $\pm$ 21 & 21 47 $\pm$ 20 & $17\pm 2$      &    4.8          & $<12.5$    &	-      &	8 &    \\
J2205+2244     & 22 05 30         & 22 44 $\pm$ 20 & $22\pm 2$      &	3.0          &	40        &	-      &	1     & \\
J2210+2118     & 22 10 07 $\pm$ 90 & 21 18 $\pm$ 20 & $45\pm 3$      &	2.6          & $<12.5$    &	-      &	3     & \\
\hline
\multicolumn{9}{l}{These are [1] \citep{Samodurov2017}, [2] \citep{Tyulbashev2018}, [3](RRATalog)}\\
\end{tabular}
\end{table*}

The Fig. 2 presents 25 RRATs out of 33 from Table 1 except for seven RRATs detected earlier in our daily monitoring and RRAT J0941+1621  (Fig. 1).

We also found 11 strong pulses for transient J1326+3346 and believe that this is a new pulsar.
The preliminary estimation period and dispersion measure for the candidate were $DM=4\pm 1$ pc/cm$^3$ and period $P=41$~ms respectively.  The coordinates of the pulsar are $\alpha_{2000}=13^h26^m42^s$, $\delta_{2000}=33^o46^\prime$ and the coordinate accuracy is the same as that of the RRATs. A full search of the signal with $0<DM<8$~pc/cm$^3$ and a period near 41 ms was the performed. We confirmed the detection of a
 new pulsar and refined the period ($P= 41.5$~ms). The dynamic spectrum of PSR~J1326+3346 and mean profile are shown in the fig.3. The time resolution in
our monitoring programme is low; therefore, details in the observed profile on the left panel cannot be identified.


 \begin{figure*}
   \centering
   \includegraphics[width=\textwidth]{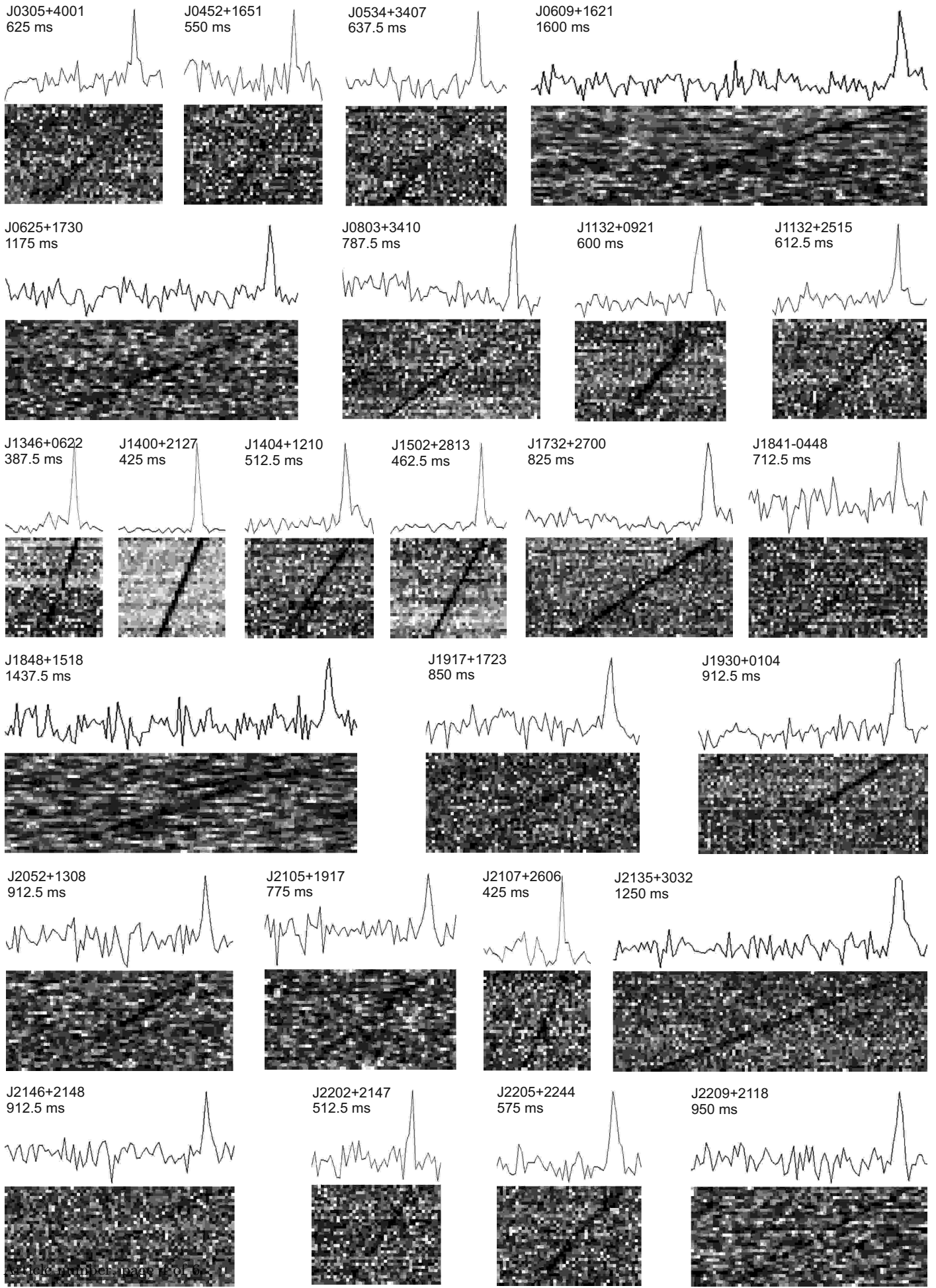}
     \caption{25 RRAT from table 1 is presented for the image. The average profile, dynamic spectrum, and recording duration are represented for each RRAT.}
   \end{figure*}

Minimum and maximum of observed dispersion measure at transients J1326+3346 ($DM = 4$~pc/cm$^3$) and J0609+1635 ($DM = 85$~pc/cm$^3$).
The distances to the new RRATs have been estimated using the NE2001 model for the electron density distribution in the Galaxy \citep{Cordes2002}
({\it https://www.nrl.navy.mil/rsd/RORF/ne2001/index.html}).
These distances do not demonstrate any peculiarities.
The median distance to the transients was determined to be $R = 1.238$~kpc (for the RRAT J2202+2147).
The nearest source is PSR J1326+3346 ($R = 0.47$~kpc), and the most distant is RRAT J2135+3032 ($R = 3.888$~kpc).

 \begin{figure}[h]
   \centering
   \includegraphics[width=\columnwidth]{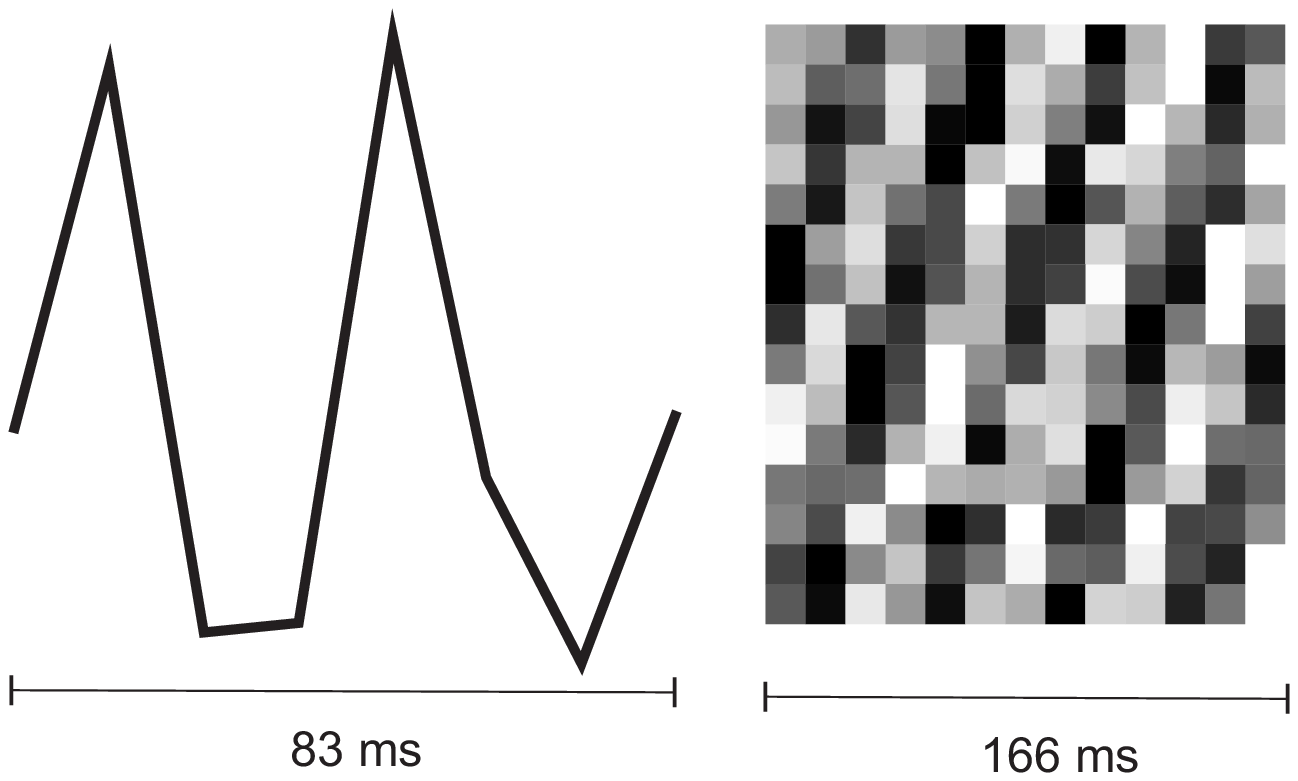}
     \caption{Dynamic spectrum of PSR J1326+3346 (right panel) has been obtained by adding with period $4\times 41.5=166$~ms for best day using programme for the search of pulsar \citep{Tyulbashev2016,Tyulbashev2017}. The mean profile with double period is presented in the left panel.}
   \end{figure}

In addition to the detected new RRATs (see Table 1 and Fig. 2) not included in the known catalogues of pulsars and transients (ATNF, RRATalog, FRBcat), we found individual pulses from 103 pulsars.
 The analysis of these pulsars will be presented in subsequent papers. The images of the dynamic spectra and short information about each pulsar are presented on the site {\it https://bsa-analytics.prao.ru/transients/pulsars.php?lang=eng }.
For comparison, ATNF catalogue contains 420 pulsars with the declinations $-9^o<\delta < 42^o$ and $DM<100$~pc/cm$^3$. Then we observed 25\% these pulsars as transients, that indicate
 the high efficiency of the pulsar search using individual pulses at a metre wavelength. The efficiency of this method of pulsar search is comparable with our method of the Fourier spectra accumulated, because we have found 131 known pulsars during three years  \citep{Tyulbashev2017}.

It should be noted that in the processed data there remain many faint RRAT candidates.
They are objects that have appeared several times; on their dynamic spectrum, the dispersion curve is virtually invisible.
Therefore, they have not been included in this work.
We may expect that with an increase of the number of processed days, additional repeated events will be observed in the same directions and at the same dispersion measures,
and the total number of RRATs can thus grow by a factor of a few.

\section{Discussion and Conclusion}

The appearance of the dynamic spectrum and the obtained mean profile of one of the detected fast transients
(J1326+3346) have suggested the existence of a new fast pulsar ($P = 41.5$~ms).
This pulsar could not be identified in the summary power spectra \citep{Tyulbashev2017}, because using this technique, the minimum possible period was $200$~ms.
The dispersion measures of all discovered transients do not contradict the NE2001 model.
In this case, we can make the conclusion that all our radio transients (Table 1) belong to the RRATs group, but not to FRB ones.

We have compared the list of the objects from RRATalog with the objects detected by our group and found coincidence for three sources.
In particular, in this catalogue, the object J0301+2052 with $DM = 19$~pc/cm$^3$ and $P = 1.207$~s is classified as a RRAT.
In our pulsar search \citep{Tyulbashev2016} we found the pulsar J0303+2248 with the same dispersion measure and period.
Seemingly, this is the same object.
A possible transition from RRAT to pulsars in observations at a lower frequency was indicated in \citet{Deneva2009}.
In RRATalog the source J1538+2345 is classified as a RRAT; however, in our observations, it has been detected on more than one hundred occasions, both as a pulsar and a RRAT.
In the ATNF catalogue, the source J1538+2345 is also listed as a pulsar.
These two sources are not included in our list (Table 1 and to figs.1-3).
The RRAT J1848+1518 ($DM=75$~pc/cm$^3$) coincides, apparently, with RRATalog source J1849+1517 ($DM=77.4$~pc/cm$^3$)
and it is the only RRAT from the full list of RRATs (RRATalog) which is included in our list and to fig.2.
We have not detected the RRAT J2225+35 found by LPA in observations of 2004 and 2006 by the group of Yu. P. Shitov \citep{Shitov2009}.
Most likely this indicates its low duty factor: one pulse in more than 10 hours (the total time of LPA observations at each sky point for a six-month period).
Considering the comparable sensitivity of the Parkes and LPA antennas (Table 1 in \citet{Tyulbashev2018}),
we may expect similar duty factors.
It should be kept in mind that among pulsars, there have been recently detected objects with very infrequent, strong bursts of pulses.
Thus, for PSR J0653+8051, two flares have been observed with a total duration of about 5 minutes during 400 days of observations;
this makes about $0.3\%$ of the total observational time, and this is equivalent to one strong pulse in 6.5 hours of continuous record \citep{Malofeev2016}.

In the paper \citet{Keane2008} an estimation was made of the possible quantity of observed RRATs (non canonical part of the full pulsar sample \citep{Karako2015}) and it is twice as many as the quantity of ordinary
(canonical) pulsars.
If the flux densities distribution of the RRATs individual pulses is the same as pulsars and, taking into
account our detection of individual pulses from 103 known pulsars,
we can estimate that only 15\% RRATs with S/N>7 that were available for observations at LPA were detected.

The main result of our rotating radio transients search at the 111 MHz frequency is the detection of 25 RRATs,
the confirmation of 8 RRATs determined from early detections, and the detection of one ms pulsar J1326+3346. Therefore
the period estimations have been made for 3 out of 25 new RRATs, and for 3 out of 8 ones detected early. We found that 16 new RRATs demonstrated more than
one events, but the remaining 9 ones were registered only once.


\begin{acknowledgements}
The authors thank L.B. Potapova and G.E. Tyul'basheva
for help with the manuscript and figures.

      This work was supported by the Russian Foundation for Basic Research (project codes 16-02-00954),
      and by the Program of the Presidium of the Russian Academy of Sciences "Transition and Explosive Processes in Astrophysics".
\end{acknowledgements}

\bibliographystyle{aa} 
\bibliography{serg} 
\end{document}